\documentclass[onecolumn,amsmath,amssymb,aps,pra,superscriptaddress,nofootinbib]{revtex4-2}

\pdfoutput=1
\usepackage[utf8]{inputenc}
\usepackage[english]{babel}
\usepackage[T1]{fontenc}
\usepackage{xcolor}
\usepackage{graphicx}
\usepackage{braket}
\usepackage{caption}
\usepackage{subcaption}
\usepackage{hyperref}
\usepackage{xr}
\usepackage{amsthm}
\usepackage{dsfont}
\usepackage{float}
\usepackage{graphicx} % Required for inserting images
\usepackage{amsmath}
\usepackage{graphicx} % Required for inserting images
\usepackage{tabularx,ragged2e}
\usepackage{array}
\usepackage{siunitx}
\usepackage{makecell}

\usepackage{graphicx}
\usepackage{booktabs}
\usepackage{multirow}
\usepackage{amssymb}
\usepackage{geometry}
\usepackage{hyperref}
\usepackage{changepage}
\usepackage{tablefootnote}
\usepackage{float}
\usepackage{xcolor}
\usepackage{tikz}
\usetikzlibrary{arrows.meta, shadows}
\usepackage{adjustbox} % Use adjustbox for tables
\usepackage{algpseudocode}
\usepackage{algorithm}
\usepackage{physics}

\usepackage[version=4]{mhchem}

\begin{document}
\title{High Performance Quantum Emulation for Chemistry Applications with Hyperion}
\author{Olivier Adjoua}\email{passi.adjoua@sorbonne-universite.fr}
\affiliation{Sorbonne Universit\'e, LCT, UMR 7616 CNRS, 75005 Paris, France}
\author{Siwar Badreddine}\email{siwar.badreddine@qubit-pharmaceuticals.com}
\affiliation{Qubit Pharmaceuticals, Advanced Research Department, 75014 Paris, France}
\author{César Feniou}
\affiliation{Qubit Pharmaceuticals, Advanced Research Department, 75014 Paris, France}
\author{Igor Chollet}
\affiliation{Université Sorbonne Paris Nord, LAGA, UMR 7539 CNRS, 93430 Villetaneuse, France}
\author{Diata Traore}
\affiliation{Qubit Pharmaceuticals, Advanced Research Department, 75014 Paris, France}
\author{Guillaume Michel}
\affiliation{Qubit Pharmaceuticals, Advanced Research Department, 75014 Paris, France}
\author{Jean-Philip Piquemal}\email{jean-philip.piquemal@sorbonne-universite.fr}
\affiliation{Sorbonne Universit\'e, LCT, UMR 7616 CNRS, 75005 Paris, France}
\affiliation{Qubit Pharmaceuticals, Advanced Research Department, 75014 Paris, France}
\date{\today}

%%%%%%%%%%% Theorems %%%%%%%
\newtheorem{thm}{Theorem}[section]
\newtheorem{prop}[thm]{Proposition}
\newtheorem{lem}[thm]{Lemma}
\newtheorem{cor}[thm]{Corollary}
\newtheorem{definition}[thm]{definition}
\newtheorem{example}{Example}[section]
\newtheorem{remark}{remark}[section]
%%%%%%%%%%%%%%%%%%%%%%%%%%%%%%%%%%%%%%%%%%
%Newcommand% symbols
%%%%%%%%%%%%%%%%%%%%%%%%%%%%%%%%%%%%%%%%%%
\newcommand{\Hyperion}{Hyperion}
\newcommand{\Hsix}{\texttt{\ce{H6}}}
\newcommand{\Hh}{\texttt{\ce{H8}}}
\newcommand{\Hdix}{\texttt{\ce{H10}}}
\newcommand{\Hdou}{\texttt{\ce{H12}}}
\newcommand{\Hq}{\texttt{\ce{H14}}}
\newcommand{\Hse}{\texttt{\ce{H16}}}
\newcommand{\Hdh}{\texttt{\ce{H18}}}

\newcommand{\vect}{\boldsymbol}
\newcommand{\Mat}{\mathbf}
\newcommand{\Ten}{\mathcal}
\newcommand{\igor}[1]{\color{orange}Igor: #1\color{black}}
\newcommand{\siwar}[1]{\color{orange}siwar: #1\color{black}}
\newcommand{\done}[1]{\color{blue}ok: #1\color{black}}

%%%%%%%%%%%%%%%%%%%%%%%%%%%%%%%%%%%%%%%%%%
%Algo
%%%%%%%%%%%%%%%%%%%%%%%%%%%%%%%%%%%%%%%%%%
\begin{abstract}
The strategic demand for quantum hardware currently outpaces the availability of near-term devices, necessitating high-performance software emulators to validate novel protocols. We introduce Hyperion, a massively parallel, GPU-accelerated quantum emulator architected to bypass the classical memory walls inherent in strongly correlated quantum chemistry simulations. Hyperion leverages custom-optimized Sparse Matrix-Sparse Vector (SpMspV) kernels to natively accelerate exact matrix-vector multiplications, enabling strictly accurate State-Vector (SV) ADAPT-VQE simulations for up to 32 qubits on multi-node platforms. To scale beyond this hardware limit, we address the  trade-off in pure Matrix Product State (MPS) emulators, where standard compression yields severe truncation errors and strict compression triggers intractable tensor rank explosions. We propose a novel partitioned emulation, namely the SV-MPS strategy: by routing non-interacting terms into an exact sparse SV core and delegating interacting terms to the MPS engine, this approach achieves emulation of 36 to 40 qubits with controlled approximations. This partitioning significantly reduces GPU resource requirements while maintaining robust accuracy across  ADAPT-VQE iterations.
Ultimately, Hyperion offers a high-fidelity platform dedicated to the development of new quantum algorithms for chemistry, enabling the modeling of realistic chemical systems at accuracies approaching the exact Full Configuration Interaction (FCI) / Complete Basis Set (CBS) limit.
\end{abstract}
\maketitle

\section{Introduction}
The rapid advancement of quantum and hybrid algorithms has created a critical demand for direct access to quantum hardware, a resource that is both strategically vital and increasingly scarce, with current availability limited primarily to Noisy Intermediate-Scale Quantum (NISQ) devices. This supply-demand imbalance has catalyzed the development of high-performance software designed to simulate quantum systems on classical architectures. These quantum emulators\cite{haner2016high,altman2021quantum} are now indispensable for validating novel protocols, refining algorithmic design, and establishing rigorous benchmarks for hardware verification.

Consequently, the last five years have seen a surge in emulation frameworks \cite{Suzuki2021,McClean2020,pennylane,Qiskit,myqlm2024,Guerreschi2020,10313722,nvidia2024cudaq}, each employing specialized strategies tailored to specific computational targets. Within this landscape, we introduce our Hyperion emulator: a GPU-accelerated, high-performance library engineered for quantum chemistry simulations. 
Although no exact emulator can permanently evade the exponential memory wall, Hyperion is engineered to  expand the accessible emulation frontier. By pushing  noiseless emulation into the 40-qubit regime, it bridges the gap between current NISQ methodologies and the algorithmic demands of the looming Fault-Tolerant Quantum Computing (FTQC) era.

By providing a controlled environment for algorithmic design, Hyperion targets chemistry problems that remain intractable for classical methods, specifically those that require the exact Full Configuration Interaction (FCI) / Complete Basis Set (CBS) limit accuracy\cite{traore2024shortcut}. While other emulators often focus on generality, Hyperion exploits symmetries and conserved quantities found in chemistry systems. It strictly uses sparse state-vector representations and  tensor networks governed by controlled approximations, ensuring high-fidelity validation of novel quantum logic.

In this paper, we detail Hyperion’s global architecture and its core emulation modules:

\begin{itemize}
    \item Hyperion-1: A dedicated state-vector (also referred to as a Schrödinger-style simulator) emulator\cite{fatima2021faster} for exact, small-to-medium scale simulations.
        \item Hyperion-2: A dual-mode emulator featuring a pure Matrix Product State (MPS)\cite{Markov2008} engine and a novel partitioned emulation SV-MPS strategy.
\end{itemize}
Within the Hyperion-2 partitioned approach, the molecular Hamiltonian is hierarchically partitioned: non-interactive local blocks are evaluated exactly using a sparse state-vector representation, while complex interactive terms are handled via compressed MPS.

We begin by introducing Hyperion's general framework.

 \section{Method}  
\subsection{Quantum Algorithms for Chemistry }
\label{seq:: general_frame}
\subsubsection{General Framework for Variational Algorithms}
In this section, we provide the generic framework for a wide variety of quantum algorithms, namely variational algorithms, especially designed for the simulation of quantum chemistry systems. 

The core objective of these algorithms is to find the ground state of a given Hamiltonian operator, denoted as  $\mathbf{H} \in \mathbb{R}^{2^n \times 2^n}$. Here, we are investigating molecular systems that possess time-reversal symmetry which yields a real Hamiltonian matrix.
For simplicity, we use the same notation for both the abstract operator and its matrix representation. The algorithm begins by preparing  a parametrized quantum state, denoted as $\psi(\boldsymbol{\theta})$, where ${\boldsymbol{\theta}= \left\{ \theta_i \right\}_{i=1}^M, M \in \mathbb{N}^*, \theta_i \in \mathbb{R}}$, is a list of parameters that completely characterize the state. The goal is  to compute the following expectation value
\begin{equation}
\label{eq::ev}
    {E}(\boldsymbol{\theta}) := \bra{\psi(\boldsymbol{\theta})}\mathcal{\Mat {H}}\ket{\psi(\boldsymbol{\theta})},
\end{equation}
with $\ket{\psi(\boldsymbol{\theta})} := \displaystyle\sum_{k=0}^{2^n-1}\omega_k(\boldsymbol{\theta})\ket{\psi_k}$ decomposed on a pure state basis $\{\ket{\psi_k}\}_k$, with $\omega_k(\boldsymbol{\theta}) \in \mathbb{C}$. The general goal of our target family of methods is to optimize $\boldsymbol{\theta}$ over $\mathbb{R}^M$  such that ${E}(\boldsymbol{\theta})$ is minimal, i.e. to find

\begin{equation}
    \boldsymbol{\theta}_{\min} := \mathop{argmin}_{\boldsymbol{\theta}\in \mathbb{R}^M}\hspace{0.05cm}{E}(\boldsymbol{\theta}).
\end{equation}
What distinguishes these families of methods is the way in which $\boldsymbol{\theta}$ is optimized, i.e., how $\ket{\psi(\boldsymbol{\theta})}$ is computed in a given iteration of the optimization procedure. Thus, each method comes with a preparation procedure  that computes $\ket{\psi(\boldsymbol{\theta})}$ from a pure state (which we assume to be $\ket{\psi_0}$ without loss of generality). This  state preparation is given by a unitary operator ${U}(\boldsymbol{\theta})\hspace{0.02cm}:\hspace{0.02cm}\mathbb{C}^{2^n}\rightarrow \mathbb{C}^{2^n}$, i.e. $\ket{\psi(\boldsymbol{\theta})} = U(\boldsymbol{\theta})\ket{\psi_0}$. The specific form of the operator $U(\boldsymbol{\theta})$ is determined by the chosen optimization method and the structure of the ansatz. Hence, Eq. \eqref{eq::ev} can be reformulated as
\begin{equation}
    {E}(\boldsymbol{\theta}) =\bra{\psi_0}U(\boldsymbol{\theta})^\dagger\Mat{H}U(\boldsymbol{\theta})\ket{\psi_0}.
\end{equation}

Evaluating the parameterized energy $E(\boldsymbol{\theta})$ requires applying a sequence of unitary transformations, typically expressed as a parameterized quantum circuit, to an initial reference state. On classical computing architectures, this state preparation and subsequent measurements can be emulated using distinct mathematical frameworks, namely state-vector formalism and tensor network approximations (low-rank representations). These foundational methods are 
detailed in Sections \ref{sec:sv_emulation}, \ref{sec:mps_emulation} and \ref{sec:sv_mps_emulation}.

Among the VQA approaches, the Variational Quantum Eigensolver (VQE)\cite{peruzzo2014variational} has been first introduced and remains the most popular hybrid quantum-classical algorithm to perform applications in quantum chemistry, quantum simulations and optimizations problems. The latter uses the variational principle to compute the ground state energy of a given Hamiltonian.
In our numerical evaluations, we select ADAPT-VQE \cite{grimsley2019adaptive}, a variant of VQE, as our primary benchmark. Widely recognized as a leading candidate for hybrid quantum-classical simulations on near-term hardware, ADAPT-VQE presents an interesting challenge for classical emulation. Unlike fixed-depth algorithms, it employs a dynamic ansatz-construction strategy, systematically expanding the electronic wave function by adding operators that most significantly lower the system's energy. While this adaptability makes it exceptionally effective for strongly correlated electronic systems where traditional methods fail, it creates a massive computational bottleneck: evaluating the energy gradient across a vast operator pool at each iteration is computationally expensive. Consequently, ADAPT-VQE serves as an ideal stress test for Hyperion. It simultaneously validates our massively parallel GPU acceleration for rapid operator evaluation and shows the necessity of the partitioned SV-MPS architecture to sustain reasonable accuracy as circuit depth and entanglement dynamically grow.

\subsubsection{Quantum Phase Estimation (QPE)}
 For the ground state quantum chemistry problem, one can encounter also Quantum Phase Estimation (QPE) algorithm\cite{lee2021even}. QPE is a fundamental quantum algorithm designed to estimate the phase $\phi$ associated with an eigenvalue $e^{2\pi i \phi}$ of a unitary operator $U$. The algorithm employs two quantum registers: a counting register to store the estimated phase and a target register containing an eigenstate of the operator. By applying a series of controlled-unitary operations followed by an inverse Quantum Fourier Transform (iQFT), the phase information is ``kicked back'' into the counting register and converted into a readable binary approximation. QPE serves as a critical subroutine for major protocols such as Shor's factoring algorithm and the calculation of ground-state molecular energies in quantum chemistry. Because the required circuit depth for high-precision estimation typically exceeds the coherence limits of current noisy hardware, QPE is primarily categorized as an algorithm for the fault-tolerant quantum computing (FTQC) era. 
 
\subsection{Overview of the Hyperion Quantum Emulator}
Hyperion is a massively parallel, GPU-accelerated quantum emulator specifically architected to fight the classical memory wall inherent in strongly correlated quantum chemistry simulations. Although its original primary objective is the  validation of novel  quantum algorithms developed within our research group, its underlying design relies on highly optimized sparse linear algebra and can be used by a larger audience. By accelerating operations between state-vectors and Hamiltonian matrices, Hyperion achieves unprecedented emulation scale and throughput. It is currently optimized for NVIDIA GPUs with CUDA and will be extended to other hardware providers in the future. In this work, we present the following core contributions:

\begin{itemize}
    \item \textbf{Massively Scalable GPU Architecture:} Hyperion is developed as a high-performance C++ library with user-friendly Python bindings, engineered specifically for multi-GPU HPC environments. We successfully deployed the emulator across 256 NVIDIA H100 GPUs spanning 64 nodes on the Jean-Zay supercomputer for a large molecular  simulation.

    \item \textbf{Novel Sparse Linear Algebra Kernels:} To maximize computational throughput during measurements, we introduce custom-optimized Sparse Matrix-Sparse Vector (SpMspV) CUDA kernels. To the best of our knowledge, Hyperion is the first quantum emulator to natively incorporate GPU-accelerated sparse matrix sparse vector (SpMspV) operations for exact matrix-vector and vector-matrix-vector multiplications, bypassing the overhead of on-the-fly operator assembly.

    \item \textbf{Exact ADAPT-VQE:} Leveraging these memory-optimized kernels, we demonstrate the capability to execute strictly exact, state-vector ADAPT-VQE simulations for up to 32 qubits, successfully running hundreds of optimization iterations without heuristic truncation.

    \item \textbf{Bounded approximation:} For larger systems, Hyperion provides a  Matrix Product States (MPS) engine. This module employs cutoff strategies to scale the number of emulated qubits while  controlling truncation errors. It uses cuTENSOR\cite{nvidia_cutensor}, cuSPARSE\cite{nvidia_cusparse} and cuSOLVER\cite{nvidia_cusolver} NVIDIA's libraries to accelerate tensor factorizations and contractions.

    \item \textbf{The Partitioned SV-MPS Emulation:} We introduce a paradigm-shifting partitioned emulation strategy that hierarchically partitions the molecular Hamiltonian. By routing non interactive terms into a sparse state-vector core (exact core), and delegating interactive terms to the MPS engine (with controlled error), we bypass the standard tensor rank explosion. This allows for near-exact emulation up to 36-40 qubits while  reducing the number of required GPU resources.
\end{itemize}

In the following sections, we introduce the theoretical and computational frameworks that underlie Hyperion's emulation engines:

\subsection{Sparse State-Vector Emulation}
\label{sec:sv_emulation}
The state-vector approach constitutes the most direct and exact method for emulating a quantum computer, as it explicitly represents the full wavefunction of an $n$-qubit system as a complex vector in a $2^n$-dimensional Hilbert space. In the context of quantum chemistry simulations, this formalism naturally accommodates the mapping between qubits and fermionic degrees of freedom, where each qubit typically encodes the occupation of a spin-orbital through standard transformations such as Jordan–Wigner\cite{jordan1928paulische} or Bravyi–Kitaev\cite{bravyi2002fermionic}. While this representation enables numerically exact simulations of quantum circuits—including entanglement and interference effects—it is inherently limited by its exponential memory footprint, scaling as $O(2^n)$, which rapidly becomes prohibitive beyond a few dozen qubits on classical hardware. Despite this constraint, Hyperion implements a state-vector simulator to validate hybrid quantum-classical algorithms such as VQE or ADAPT-VQE, where its attributes provide noiseless reference implementations of parameterized circuits and allow precise evaluation of expectation values of molecular Hamiltonians. 
%This makes them indispensable for benchmarking ansätze, studying convergence properties, and guiding the design of more resource-efficient quantum algorithms for ground-state energy estimation.

While most quantum circuit simulators rely on dense state-vector representations (see NVIDIA CUDA-Q for a representative example \cite{nvidia2024cudaq}), alternative approaches inspired by quantum chemistry and exact diagonalization techniques have explored sparse formulations of the Hamiltonian in a second-quantized basis\cite{knowles1984new}. In this spirit, Hyperion  departs from standard gate-based emulation by explicitly assembling the molecular Hamiltonian on GPU architectures and storing it in a distributed compressed sparse row (CSR) format, obtained by summing the contributions of individual fermionic operators after qubit mapping. Although such sparse Hamiltonian constructions are common in classical Full Configuration Interaction (Full CI) solvers, they are less frequently employed in quantum circuit emulators, which typically favor operator-based (matrix-free) applications to avoid materializing exponentially large objects.

In contrast, Hyperion leverages the intrinsic structure of quantum chemistry problems by restricting the Hilbert space to physically relevant subspaces. The state-vector $\ket{\psi}$ is therefore also stored in a sparse format, which is particularly well suited when working within the Full CI manifold ($\Omega_{CI}$ ), whose cardinality scales combinatorially as the number of ways to distribute $Ne$ electrons among $N$ spin-orbitals (with $Ne< N$), rather than exponentially as in the full Hilbert space ($\Omega_H$). Furthermore, by enforcing global symmetries such as spin conservation, the accessible configuration space can be reduced to a smaller subspace $\Omega_{CI_k}\subset \Omega_{CI}$, yielding a substantial reduction in both memory footprint and computational cost, see Tab.\ref{tab:hilbert_ci_sizes}. This symmetry-adapted construction enables the Hamiltonian to be directly assembled within $\Omega_{CIk}$, significantly reducing its effective dimension.

In practical regimes relevant to molecular systems (e.g., hydrogen chains), the resulting state-vector exhibits a sparsity pattern with a filling ratio typically below 5\% in the worst case. One can refer to column 7 of Tab.\ref{tab:hilbert_ci_sizes}. Combined with an initial state $\ket{\psi_0}$ corresponding to the Hartree–Fock configuration—often itself a single determinant — this justifies the use of sparse linear algebra as the computational backbone of the simulator. This design choice positions Hyperion at the interface between quantum circuit emulation and classical electronic structure methods, while maintaining compatibility with hybrid algorithms such as VQE, where repeated Hamiltonian applications and expectation value evaluations benefit directly from sparse representations.

The primary challenge associated with this approach lies in the efficient implementation of the linear algebra operations involving the Hamiltonian and the state-vector $\ket{\psi}$. Standard high-performance libraries such as Intel MKL, BLAS, or NVIDIA cuSPARSE\cite{nvidia_cusparse} provide highly optimized kernels for operations involving sparse matrices and dense vectors, but offer limited or no support for sparse–sparse primitives, particularly in the context of GPU acceleration. Since the central design objective of Hyperion is to minimize memory footprint without compromising computational performance, relying on dense intermediates is not a viable option. This constraint motivated the development of a dedicated GPU-oriented and optimized library implementing sparse–sparse linear algebra operations, writen in CUDA C/C++, and tailored to our use case. These operations include, in particular, matrix–vector products between a CSR Hamiltonian and a sparse state-vector, inner products between sparse vectors, and the application of exponentiated fermionic operators to sparse states, yielding sparse outputs while preserving the structure of the wavefunction.

From a parallelization perspective, the design philosophy minimizes inter-GPUs communications while pushing further up the number of qubits that can be simulated. However, distributing sparse data structures introduces nontrivial communication challenges. In the worst case, a sparse state-vector distributed over $P$ MPI processes may induce up to $O(P^2)$ communication patterns during Hamiltonian application, depending on the sparsity structure and data layout, thus severely degrading performance. To mitigate this issue, we adopt a strategy in which only the Hamiltonian is distributed across processes, with its rows uniformly partitioned according to the underlying symmetry-restricted subspace $(\Omega_{CIk})$. The state-vector, by contrast, is fully replicated on each MPI process. During computation, each process evaluates only the subset of vector components corresponding to its assigned Hamiltonian rows, and collective communication is limited to a single all-to-all reduction and/or broadcast step at the end of each operation.

This design significantly reduces communication overhead while maintaining a controlled memory footprint, ultimately enabling the emulator to leverage large-scale classical supercomputing resources for electronic structure calculations. As a result, Hyperion achieves an effective compromise between memory efficiency, parallel performance, and algorithmic flexibility, allowing the simulation of quantum algorithms within physically relevant subspaces that would otherwise be inaccessible using conventional dense state-vector approaches for high qubit number.
In other words, Hyperion offers the possibility to efficiently compute hundreds if not thousands of ADAPT-VQE iterations on relevant molecular systems, see Figures \ref{fig:small_chains} \ref{fig:big_chains}, an essential asset to enable the study of adaptive methods. Table \ref{tab:hilbert_ci_sizes} displays an example of the performances of the SV emulator using hydrogen chains up to 16 atoms, i.e. 32 qubits using the STO-3G basis set.

 \subsubsection{Numerical performance}
A particularly relevant algorithm for assessing the capabilities of a state-vector-based quantum emulator is ADAPT-VQE. This hybrid quantum-classical method constructs the variational ansatz iteratively by selecting, at each step, the operator from a predefined pool (e.g., qubit excitation-based, QEB) that maximally reduces the energy gradient. The ansatz thus grows dynamically, with the number of variational parameters $\mathbf{\theta}$ increasing at each iteration, followed by a classical optimization phase. Convergence is typically assessed through the norm of the gradient or the variation of the energy expectation value. This iterative structure makes ADAPT-VQE particularly suitable for probing the asymptotic behavior of quantum simulators, as both the circuit depth and the complexity of state preparation increase with the number of iterations.

Within Hyperion, the ability to efficiently manipulate sparse state-vectors enables the exploration of increasingly large qubit systems while maintaining a tractable memory footprint. For performance evaluation and validation, we consider linear hydrogen chains ranging from \ce{H4} to \ce{H16}, providing a systematic increase in qubit count, respectively 8 to 32 qubits. Molecular integrals are computed using the PySCF package, and simulations are initialized from the Hartree–Fock reference state. These systems offer a convenient and scalable benchmark, as the size of the underlying Hilbert and configuration spaces grows rapidly with system size. All simulations were performed on the Jean Zay supercomputer, specifically on the H100 GPU partition, which features NVIDIA Hopper GPUs interconnected through high-bandwidth NVLink and supported by a high-performance InfiniBand network, providing a suitable environment for large-scale distributed sparse computations.

\begin{figure*}[tbh]
    \centering
    \includegraphics[width=\textwidth]{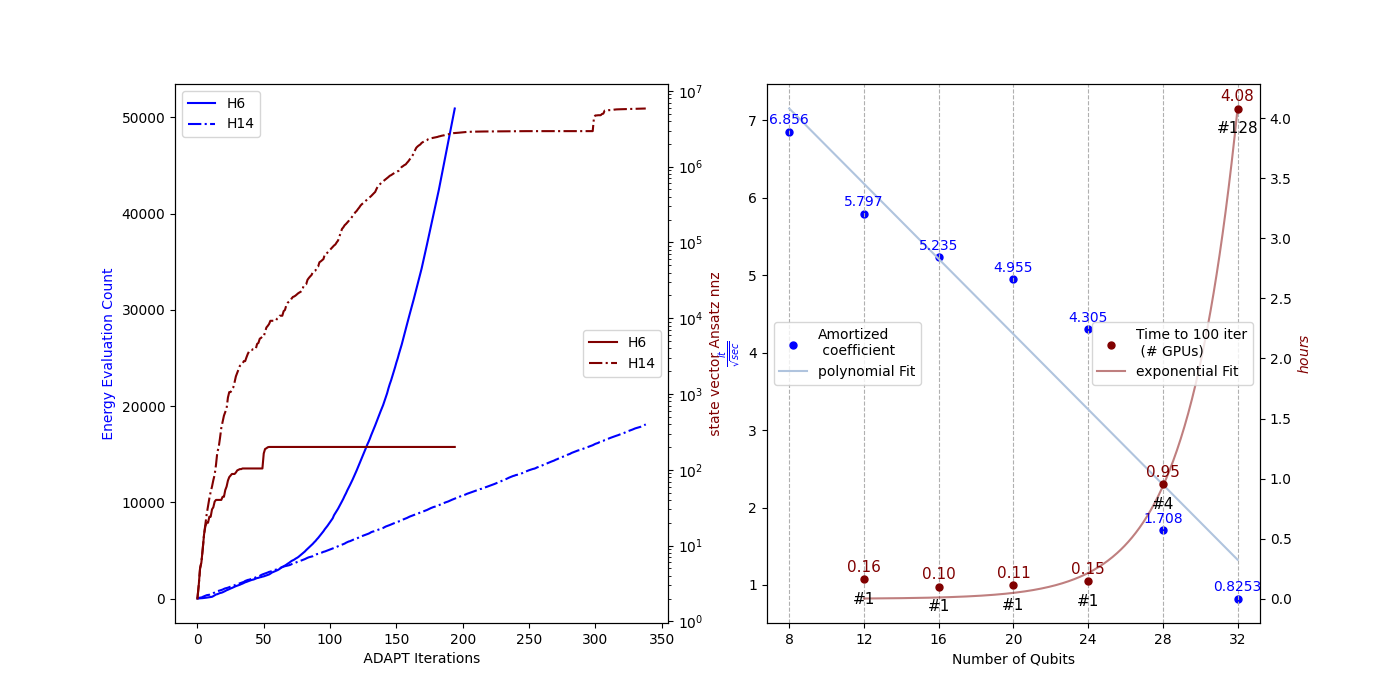}
    \caption{Algorithmic behavior and performance metrics of ADAPT-VQE simulations using Hyperion on hydrogen chains.
    Evolution of ADAPT-VQE statistics for \ce{H6} and \ce{H14} as a function of the iteration index on left. The left axis shows the cumulative number of energy evaluations (expectation values), highlighting two distinct regimes for \ce{H6}: an initial linear growth up to $\sim$ 70 iterations, followed by a polynomial regime associated with increased optimization cost. For $H14$, only the linear regime is observed within the simulated range (up to 340 iterations). The right axis reports the number of nonzero components in the sparse state-vector, illustrating the progressive growth of the ansatz, interspersed with plateaus corresponding to optimization phases. For \ce{H6}, saturation occurs at approximately 50\% of the symmetry-restricted configuration space $\Omega_{CIk}$, coinciding with the onset of the polynomial regime. A similar trend is expected for \ce{H14}, although the asymptotic regime is not reached within the considered iterations. }
    \label{fig:HypPerf}
\end{figure*}

The results are summarized in Fig.\ref{fig:HypPerf}  which presents two complementary analyses. The left panel reports algorithmic statistics for representative systems (\ce{H6} and \ce{H14}). The number of energy evaluations (i.e., expectation value computations) exhibits two distinct growth regimes for \ce{H6}: an initial linear regime up to approximately \num{70} iterations, followed by a polynomial regime associated with increased optimization cost as the ansatz becomes more expressive. For \ce{H14}, only the linear regime is observed within the simulated iteration range (up to \num{340} iterations), indicating that the asymptotic regime has not yet been reached. Simultaneously, the number of nonzero elements in the sparse state-vector increases as the ansatz grows, with intermittent plateaus corresponding to optimization phases. For \ce{H6}, saturation occurs around 50\% of the symmetry-restricted configuration space $\Omega_{CIk}$, coinciding with the transition to the polynomial regime. A similar trend is expected for larger systems, although not fully observed due to computational limits.

To quantify performance in a hardware-agnostic manner, we introduce a metric derived from the early (linear) regime of ADAPT-VQE. Restricting the analysis to this regime is justified both by its clear identification across all systems and by the absence of sufficient iterations to reach the polynomial regime for larger molecules (\ce{H14}, \ce{H16}). Let $\jmath$ denote the iteration index. Empirically, the walltime required to reach iteration $\jmath$ follows a linear scaling: $$ T_\jmath = c \cdot \jmath, $$
where $c$ is an effective cost per iteration. However, each iteration involves the evaluation of the energy, which itself depends on the size of the ansatz (i.e., $\jmath$). Decomposing this cost yields: $$ c = T_O\cdot \jmath + T_1, $$
where $T_0$ represents the cost of applying exponentiated fermionic operators (state preparation), and $T_1$ corresponds to the evaluation of the expectation value $ E(\boldsymbol{\theta}) $. Substituting, one obtains:
\begin{equation*}
\begin{split}
    T_\jmath = T_0\jmath^2 + T_1j \;&\approx\; C\jmath^2 , \\
              T_j &\approx C\jmath^2  \quad \Rightarrow\quad \jmath = \tilde{C} \sqrt{T_\jmath} 
\end{split}
\end{equation*}
This quadratic scaling motivates the definition of a performance metric:
\begin{equation}
    \tilde{C} = \frac \jmath {\sqrt{T_\jmath}},
\end{equation}
which can be interpreted as an ADAPT amortized iteration coefficient, homogeneous to the inverse square root of time. Importantly, the gradient evaluation step remains approximately constant throughout the simulation due to the fixed operator pool, and therefore does not significantly impact this metric.

The right panel of Fig.\ref{fig:HypPerf} reports the evolution of this coefficient as a function of the number of qubits, revealing an approximately linear decrease, as confirmed by a fitted trend. This behavior provides a meaningful proxy for the scalability of the simulator with system size. In contrast, the walltime required to perform a fixed number of ADAPT iterations (e.g., \num{100}) exhibits an exponential increase both with the number of qubits and the number of GPUs allocated, reflecting the intrinsic complexity of state-vector-based approaches despite the use of sparse representations. The \ce{H16} example speaks for itself as the Hamiltonian size imposes a total of 128 GPUs to be accessible, while only 4 GPUs is necessary for \ce{H14}.

Overall, these results highlight both the strengths and limitations of the proposed approach: while sparse representations and symmetry restrictions enable simulations at larger scales than dense methods, the exponential nature of the underlying Hilbert space remains the fundamental bottleneck, emphasizing the importance of algorithmic and architectural optimizations.

\begin{table*}[htb]
\centering
\begin{tabular}{|c|c|c|c|l|l|c|c|}
\hline
Hydrogen chains & $N_e$ & $N$ qubits & $|\Omega_H|=2^N$ & $|\Omega_{CI}|= \binom{N}{N_e}$ & $|\Omega_{CI_k}|$ & $\frac{|\Omega_{CI_k}|}{|\Omega_H|}$ & Hamiltonian Size (GB) \\
\hline
 \ce{H6} & 6 & 12 & $2^{12}$ & \num{924} & \num{400} & \num{0.098} & \num{2.03e-3}\\
 \hline
 \ce{H8} & 8 & 16 & $2^{16}$ & \num{12870} & \num{4900} & \num{0.075} & \num{1.08e-2}\\
 \hline
 \ce{H10} & 10 & 20 & $2^{20}$ & \num{184756} & \num{63504} & \num{0.061} & \num{3.31e-1} \\
\hline
 \ce{H12} & 12 & 24 & $2^{24}$ & \num{2704156} & \num{853776} & \num{0.051} & \num{5.97} \\
\hline
 \ce{H14} & 14 & 28 & $2^{28}$ & \num{40116600} & \num{11778624} & \num{0.044} & \num{226.31} \\
\hline
 \ce{H16} & 16 & 32 & $2^{32}$ & \num{601080390} & \num{165636900} & \num{0.039} & \num{7.224e3} \\
\hline
\end{tabular}
\caption{Table presenting the number of electrons, number of qubits, and the sizes of the Hilbert, Full CI, and spin-constrained Full CI subspaces ($S=0$) for hydrogen chains.}
\label{tab:hilbert_ci_sizes}
\end{table*}

\subsubsection{ Numerical Results }

\begin{figure*}[htb]
    \centering
    \begin{subfigure}[b]{0.48\textwidth}
    \includegraphics[width=\linewidth]{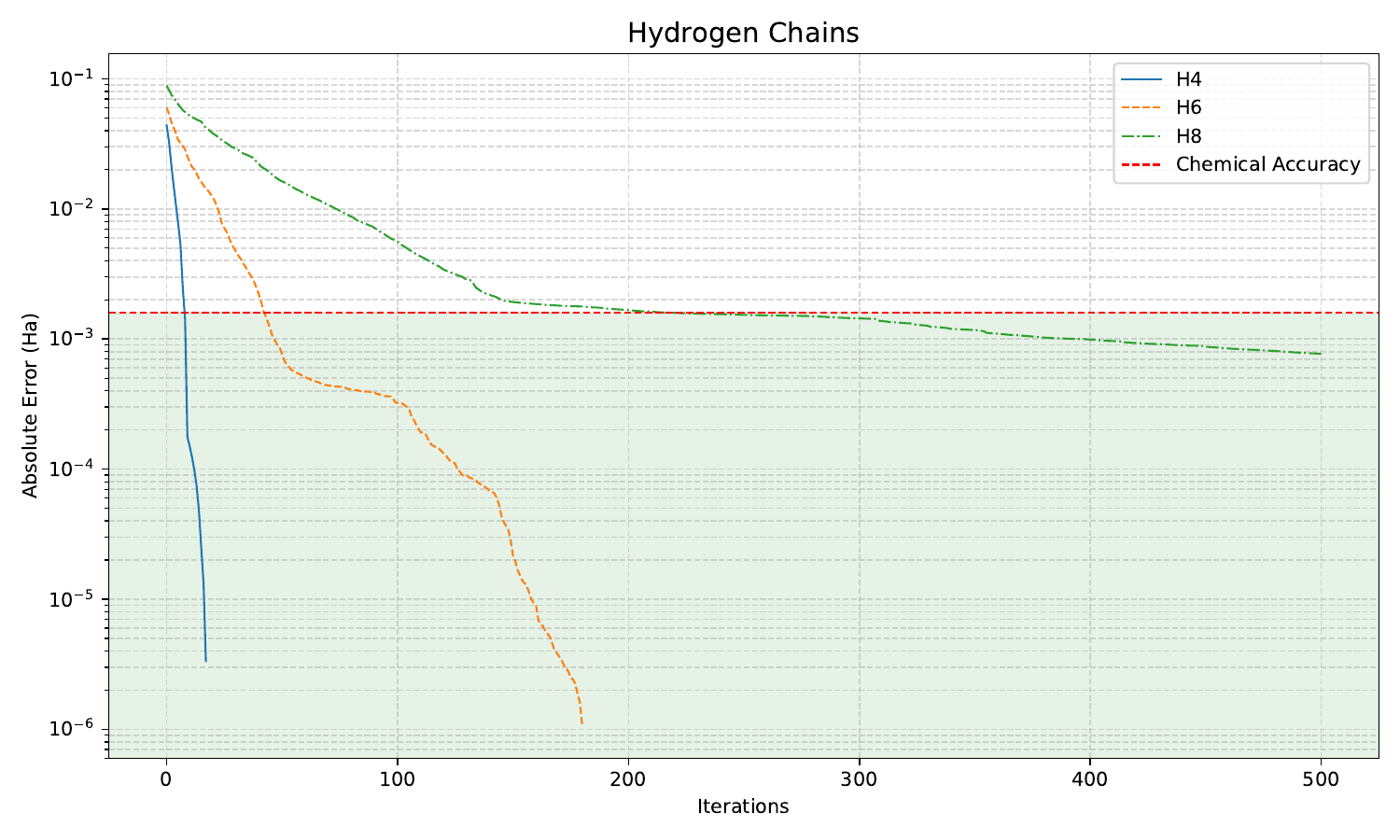}
    \caption{Energy residual error evolution across ADAPT-VQE iterations for \ce{H4}, \ce{H6} and \ce{H8} in state-vector mode. Small systems (\ce{H4}, \ce{H6}, \ce{H8}), corresponding to 8, 12, and 16 qubits, respectively. The shaded region below \SI{2e-3}{Ha} indicates the domain of chemical accuracy. \ce{H4} converges rapidly to \SI{2e-5}{Ha} within 19 iterations. \ce{H6} reaches \SI{1e-6}{Ha} after 194 iterations ($\approx 2$ hours of simulation), while \ce{H8} attains \SI{8e-6}{Ha} after \num{523} iterations ($\approx 20$ hours).}
    \label{fig:small_chains}
    \end{subfigure}
    \hfill
    \begin{subfigure}[b]{0.48\textwidth}
    \includegraphics[width=\linewidth]{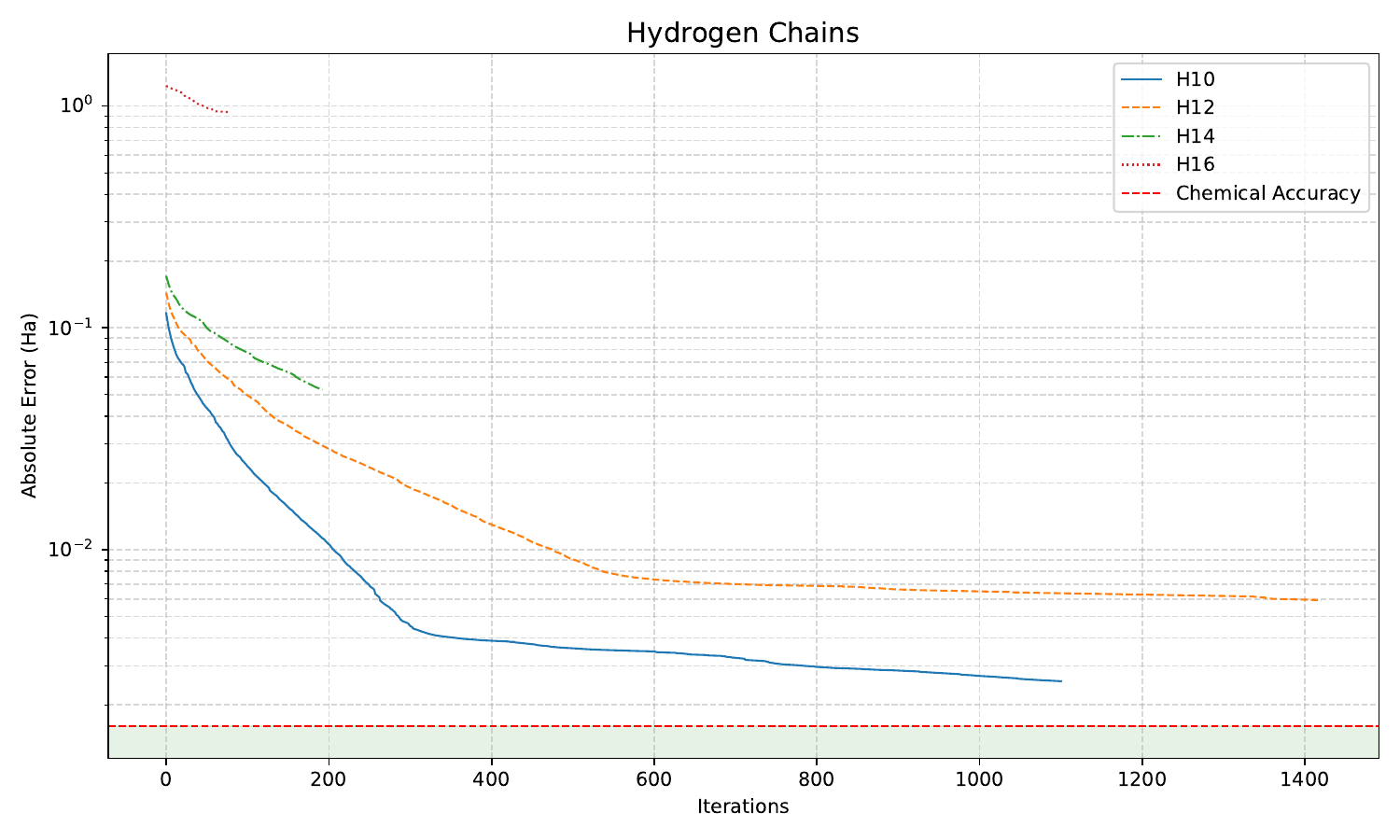}
    \caption{Larger systems (\ce{H10}, \ce{H12}, \ce{H14}, \ce{H16}), corresponding to 20, 24, 28, and 32 qubits. None of these systems reach chemical accuracy within the simulated iteration range. \ce{H10} performs \num{1116} iterations, reaching \SI{3e-3}{Ha} after 4 days; \ce{H12} reaches \SI{8e-3}{Ha} after \num{1425} iterations (4 days); \ce{H14} reaches \SI{6e-2}{Ha} after \num{340} iterations (12 hours on 4 GPUs); and \ce{H16} reaches \SI{2e-2}{Ha} after \num{80} iterations (2 hours on \num{128} GPUs). The chemical accuracy threshold is also indicated.}
    \label{fig:big_chains}
    \end{subfigure}

    \caption{Absolute energy error (Hartree) as a function of ADAPT-VQE iterations leveraging state-vector for hydrogen chains, shown on a semi-logarithmic scale. The green shaded region indicates chemical accuracy.}
    \label{fig:Hchain_convergence}
\end{figure*}

%Qualitative Analysis of ADAPT-VQE Convergence
A qualitative assessment of the ADAPT-VQE simulations across hydrogen chains reveals a clear dichotomy between small and larger systems when using a state-vector based emulator. As depicted in Fig.\ref{fig:small_chains}, for the smallest molecules (\ce{H4} and \ce{H6}), the algorithm demonstrates robust convergence toward the ground-state energy, reaching well below chemical accuracy within a moderate number of iterations and computational time. These results validate both the correctness of the implementation and the effectiveness of the sparse state-vector approach in regimes where the configuration space remains tractable. For \ce{H8}, the convergence profile exhibits a noticeable inflection point, marking a transition from the initial linear regime to the polynomial regime previously identified in the performance analysis. This transition coincides with a saturation of the number of configurations contributing to the ansatz, suggesting that the variational space explored by the algorithm approaches a significant fraction of the symmetry-restricted subspace $\Omega_{CIk}$. As a result, the optimization of the variational parameters $\boldsymbol{\theta}$ becomes increasingly costly and less effective at reducing the energy error.

This behavior may be indicative of phenomena related to optimization landscape flattening, sometimes associated with barren plateaus in variational quantum algorithms. However, strictly speaking, barren plateaus \cite{McClean_2018,Zhao_2021,larocca2025barren} refer to an exponential suppression of gradient magnitudes with system size, typically observed in deep, unstructured ansätze. In the present case, the ADAPT-VQE construction—being adaptive and operator-driven—mitigates this effect to some extent. Therefore, the observed slowdown is more accurately attributed to the growing redundancy and reduced expressibility gain of newly added operators within an already large ansatz, rather than a genuine barren plateau regime.

All systems in \ref{fig:small_chains} ultimately reach chemical accuracy within reasonable computational effort, confirming the suitability of ADAPT-VQE for small-to-intermediate problem sizes. In contrast in Fig.\ref{fig:big_chains}, none of the larger systems (\ce{H10}–\ce{H16}) achieve chemical accuracy within the explored iteration budgets. Nevertheless, these simulations are pushed to iteration counts rarely reported in the literature, providing valuable insight into the asymptotic behavior of the algorithm. As expected, the transition toward the polynomial regime leads to a strong increase in the amortization cost per iteration, with the classical optimization of the parameters rapidly becoming the dominant computational bottleneck.

Furthermore, as the gradient magnitude decreases during the optimization process, the numerical precision required to reliably update the parameters must be correspondingly higher, further increasing computational cost. Although the polynomial regime is not fully reached for the largest systems (\ce{H14}, \ce{H16}), the observed trends and simulation times already indicate a steep escalation in resource requirements.

These results highlight two fundamental limitations. First, an algorithmic limitation inherent to ADAPT-VQE, where the growth of the ansatz leads to diminishing returns in energy minimization. Second, a resource limitation associated with state-vector-based emulation, where even optimized sparse representations and large-scale GPU parallelism are insufficient to scale efficiently beyond a certain number of qubits (here, around 32, with projections suggesting significant challenges beyond $\sim$ 36 qubits). Together, these observations underscore the need for both improved ansatz construction strategies and more scalable simulation paradigms.

%transition 
It must be emphasized that SV methods are  bounded by the exponential growth of the Hilbert space. While leveraging sparsity and symmetry-adapted subspaces significantly reduces the computational prefactor, it does not alter the asymptotic scaling. Consequently, the  memory overhead required to store the sparse Hamiltonian matrix consistently drives the simulation into a hard hardware limit for larger systems beyond $32$ qubits.
A highly promising avenue for overcoming this \textit{curse of dimensionality} lies in Tensor Network (TN) methods \cite{Markov2008,Vidal2003,Schollwck2011,Cirac2021}. In this paradigm, a quantum circuit is cast as a network of low-dimensional arrays, enabling the extraction of specific amplitudes and expectation values through targeted pairwise contractions. 

Within this context, Hyperion is engineered as a dual-mode platform,  supporting both exact sparse SV and approximate MPS features. Because both modalities rely heavily on massive contraction operations, Hyperion's hardware scalability is strictly dictated by the optimization of its core linear and multilinear algebra kernels. To contextualize this architecture, the following section establishes the theoretical framework of compressed tensor networks and their specific integration into Variational Quantum Algorithms (VQAs).

%\begin{figure}[htbp]
%    \centering
%    \includegraphics[width=0.4\textwidth]{small_chains_iter_timings.pdf}
%    
%    \caption{Iteration time across ADAPT-VQE iterations for \ce{H4}, \ce{H6} and \ce{H8} in state-vector mode.}
%    \label{fig:small_chains_iter_timings}
%\end{figure}

% \begin{figure}[htbp]
%     \centering
%     \includegraphics[width=0.4\textwidth]{big_chains_iter_timings.pdf}
    
%     \caption{Iteration time across ADAPT-VQE iterations for \ce{H10}, \ce{H12} and \ce{H14} in statev-ector mode.}
%     \label{fig:big_chains_iter_timings}
% \end{figure}

\subsection{\textbf{Compressed Tensor Networks (MPS)}}
\label{sec:mps_emulation}
\subsubsection{Theoretical background}

Within the broader TN paradigm, specific topological decompositions such as Matrix Product States (MPS) and Matrix Product Operators (MPO), mathematically formalized as Tensor Train (TT) decompositions \cite{Markov2008,Vidal2003,Schollwck2011,Cirac2021,Oseledets2011}, introduce controlled approximations to achieve massive data compression. These data structures originally appeared in the Density Matrix Renormalization Group
(DMRG) algorithm ~\cite{White1992,DMRG,Schollwck2011}, which is a popular optimization technique to approximate the ground-state energy of strongly correlated quantum systems.

For quantum systems exhibiting bounded entanglement, an MPS encapsulates the  state vector within a 1D chain of low-rank tensors. This  representation shifts the memory footprint from an exponential to a polynomial scaling regime, providing a  flexible and memory-efficient framework for large-scale quantum emulation.

A general $n$-qubit state $|\psi\rangle \in \mathbb{C}^{2^n}$ can be expressed as a superposition over the computational basis:
\begin{equation}
    |\psi\rangle = \sum_{i_1,\ldots,i_n \in \{0,1\}} \Psi_{i_1\ldots i_n} |i_1\rangle \otimes \ldots \otimes |i_n\rangle,
\end{equation}
where the tensor $\Psi_{i_1\ldots i_n} \in \mathbb{C}$ contains the complex amplitudes of the  state. To mitigate the exponential scaling of this  vector, the state can be recast into a MPS framework. In this representation, each  amplitude is  factorized into a sequence of local matrix multiplications:

$$\Psi_{i_1\ldots i_n} = \mathbf{A}_1[i_1] \mathbf{A}_2[i_2] \ldots \mathbf{A}_{n}[i_{n}],$$

where $(\Ten{A}_1,\ldots,\Ten{A}_{n})$ is the set of TT-cores (local tensors), where $\Ten{A}_k \in  \mathbb{R}^{r_{k-1} \times 2 \times r_k}$ ($r_0=r_{n}=1$), and $r_k$ are referred to as the bond dimensions (or also known as TT-ranks) and ${\Mat{A}_k[i_k]:=\Ten{A}_k[:,i_k,:] \in \mathbb{R}^{r_{k-1} \times r_k}}$ for $k \in \{1,\ldots, n\}$. The bond dimensions can be thought of as a parameter controlling the expressivity of a MPS/TT network.

In a similar fashion, the Hamiltonian $\Mat{H}$ can be  factorized into a 1D chain of local tensors The TT representation of an operator is known as a Matrix Product Operator (MPO). The MPO expresses the tensor operator as a contraction product of third-order and fourth-order TT-cores, For a detailed mathematical derivation of these structures and associated arithmetics operations, we refer the reader to \cite{Schollwck2011}.

\subsubsection{MPS based VQE}

Variational Quantum Eigensolvers (VQEs) operate by optimizing a parameterized trial state $|\psi(\boldsymbol{\theta})\rangle = U(\boldsymbol{\theta})|\psi_0\rangle$, where $|\psi_0\rangle$ represents the initial reference state and $\boldsymbol{\theta} = (\theta_1, \ldots, \theta_k)$ denotes the variational parameters. The  unitary operator $U(\boldsymbol{\theta})$ is constructed as a sequential product of local unitaries, $\prod_{j=1}^k U_j(\theta_j)$. To integrate this formalism into a tensor network paradigm, these unitary transformations are mapped to exact MPO representations, which in our implementation exhibit small, strictly bounded bond dimensions ($1 \le R \le 5$). During the optimization loop, a classical routine iteratively updates $\boldsymbol{\theta}$ by evaluating expectation values with respect to an observable, typically the molecular Hamiltonian. Analogous to the sparse state-vector methodology, this Hamiltonian is explicitly assembled during the precomputational phase. By summing the mapped fermionic operators on the GPU and enforcing a  TT-rounding threshold of $\delta = 10^{-14}$, the full Hamiltonian is stored as a highly compressed, static MPO prior to initiating the VQE loop.

According to standard tensor arithmetic \cite{Schollwck2011,Lee2017}, applying a rank-1 MPO to a rank-1 MPS leaves the internal bond dimension of the target MPS invariant, whereas higher-rank MPOs multiplicatively increase it. To prevent memory exhaustion, the network must be compressed via a sequence of truncated Singular Value Decompositions (SVDs) \cite{Horn1985}, known as TT-rounding \cite{Oseledets2011}. This procedure reduces bond dimensions of tensor trains  while controlling the approximation error per compression, bounded by $\epsilon = \frac{\delta\sqrt{n-1}}{||\mathbf{A}||_F}$, where $\delta$ is the truncation threshold, $n$ is the qubit count, and $||\mathbf{A}||_F$ is the Frobenius norm of the uncompressed original input tensor. A comprehensive derivation of this error bound is  given in \cite{Oseledets2011}.

Within our ADAPT-VQE implementation, TT-rounding is  used to overcome the exponential bond dimension growth. The computational complexity of this compression is governed by three primary parameters: the total number of qubits $n$, the maximum  bond dimension  of the ansatz MPS $r$, and the bond dimension of the Hamiltonian MPO $R$. Consequently, the asymptotic scaling for the core algorithmic operations evaluates as follows:
\begin{itemize}
    \item \textbf{Gradient evaluation} across an operator pool of size $M$ scales as $\mathcal{O}(Mnr^2R(r+R))$, where we employ the "zip-up" method by Stoudenmire and White \cite{Stoudenmire2010} to suppress exponential rank growth during intermediate contractions.
   
\item \textbf{Ansatz Optimization:} Employing gradient-free classical optimizers necessitates the sequential application of $k$ exponentiated candidate operators. Because this evolution requires intermittent TT-rounding, both the ansatz optimization and the initial state preparation scale asymptotically as $\mathcal{O}(nkr^3)$. When combined with the expectation value evaluation (used for classical optimizer), which demands $\mathcal{O}(n(r^3R+r^2R))$; the total computational complexity for the optimization phase evaluates to $\mathcal{O}(n(kr^3+r^3R+r^2R))$.
    
\end{itemize}

Although these operations scale linearly with $n$, the  polynomial scaling with respect to the bond dimensions highlights the necessity of carefully tuned truncation thresholds to avoid high computational costs. To overcome the exponential memory wall of exact SV simulations without succumbing to the severe truncation errors inherent in  MPS, we propose a novel Partitioned SV-MPS emulation technique. This novel approach is detailed in the following section.

\subsubsection{Partitioned SV-MPS Emulation}
\label{sec:sv_mps_emulation}
Hyperion-2 introduces a novel partitioned SV-MPS emulation where the molecular Hamiltonian undergoes a hierarchical left-right decomposition. This structural partitioning explicitly isolates the operator's components: non-interacting local blocks are evaluated exactly via a sparse SV formalism, whereas the  interacting terms are approximated through a compressed MPO.
For a specified partitioning level $\eta \in \mathbb{Z}^+$, this decomposition takes the following general form:
\begin{equation}
\Mat{H}=\Mat{H}_{\eta}+\sum_{i=1}^{\eta} \sum_{\ell=1}^{2^{\eta-1}}\Mat{B}_{i,\ell},
\end{equation}
where $\Mat{H}_{\eta}$ is defined as:
\begin{equation}
\Mat{H}_{\eta}=\sum_{i=1}^{2^\eta} \bigotimes_{j=1}^{\frac{n(i-1)}{2^\eta}}\Mat{I}_2 \otimes \Mat{H}^{(i)} \otimes \bigotimes_{j=\frac{ni}{2^{\eta}}+1}^{n}\Mat{I}_2,
\end{equation}
with each  term $\Mat{H}^{(i)} \in \mathbb{R}^{2^{\frac{n}{2^\eta}} \times 2^{\frac{n}{2^\eta}}}$.  
The remaining terms $\Mat{B}_{i,\ell}$ capture interactions across different partitions and hierarchical levels.

By employing this partitioned formulation of the $n$-qubit Hamiltonian ($\mathbf{H} \in \mathbb{R}^{2^n \times 2^n}$), Hyperion-2 uses this formalism to execute the heaviest workloads in ADAPT-VQE: 
gradient measurements and classical energy evaluations. 
This is achieved by explicitly representing the trial ansatz into an exact sparse SV core, denoted as $\Psi_{SV}$ and a compressed MPS, denoted as $\Psi_{MPS}$. Consequently, all expectation values are computed additively, combining exact local evaluations with approximate MPS contractions over the interacting terms. For example, an expectation value with respect to the partitionned Hamiltonian can be evaluated as follows:

\begin{equation}
       E(\boldsymbol{\theta})=  \langle\Psi|\Mat{H}|\Psi\rangle =  \langle\Psi_{\text{SV}}|\Mat{H}_{\eta}|\Psi_{\text{SV}}\rangle +\sum_{i=1}^\eta \sum_{\ell=1}^{2^\eta-1} \langle\Psi_{\text{MPS}}|\Mat{B}_{i,\ell}|\Psi_{\text{MPS}}\rangle,
\end{equation}

By employing this partitioned methodology, we eliminate the necessity of storing the sparse Hamiltonian in memory. Instead, exact storage is  confined to the non-interacting local blocks, which scale efficiently as $\mathcal{O}(2^{\frac{n}{2^\eta}})$, where $\eta$ denotes the partitioning level. Conversely, the interacting Hamiltonian terms are compressed into a low-rank MPO format, with their corresponding evaluations executed via efficient tensor network arithmetic. 

\begin{figure*}[htb]
    \centering
\begin{tikzpicture}[font=\sffamily, >=Stealth]

    % --- 1. Background Grid & Axes ---
    % Y-Axis (System Size)
    \draw[->, thick] (0,0) -- (0, 8) node[above, font=\bfseries] {System Size (Number of Qubits)};
    % X-Axis (Correlation Strength)
    \draw[->, thick] (0,0) -- (11, 0) node[right, font=\bfseries] {Correlation};
    
    % X-Axis Labels
    \node[below] at (2.5, -0.2) {Weak Correlation };
    \node[below] at (8, -0.2) {Strong Correlation};

    % Y-Axis Ticks
    \draw[thick] (-0.15, 3.5) -- (0.15, 3.5) node[left=0.2cm, font=\bfseries] {~32};
    \draw[thick] (-0.15, 5.5) -- (0.15, 5.5) node[left=0.2cm, font=\bfseries] {~40};

        % --- 2. The SV Foundation (Exact Calculation) ---
        \shade[top color=orange!40, bottom color=orange!60, draw=orange!80, thick] 
            (0.05, 0.05) rectangle (10.5, 3.5);
        \node[align=center, font=\normalsize\bfseries, text=black!80] at (5.25, 2) {Hyperion-1 (State-Vector)};
        \node[align=center, font=\small, text=black!80] at (5.25, 1.1) { Exact Evaluation \\ Zero Truncation Error \\ \vspace{2pt} \textbf{Hardware Cost (32q): $\sim$128 GPUs}};

    % The Memory Wall Line
    \draw[red, ultra thick, dashed] (0, 3.5) -- (10.5, 3.5);
    \node[red!80!black, above right, font=\small\bfseries] at (10.5, 3.5) {};

    % --- 3. The Pure MPS Regime (Left side, Weak Correlation) ---
    % Survives past 32 qubits, but only for weak correlation.
    \shade[top color=cyan!20, bottom color=cyan!40, draw=cyan!80, thick] 
        (0.05, 3.5) rectangle (5, 7.5);
    \node[align=center, font=\large\bfseries, text=black!80] at (2.5, 6) {Pure MPS};
    \node[align=center, font=\small, text=black!80] at (2.5, 5.3) {Global approximation};

    % --- 4. The MPS Failure Zone (Right side, Strong Correlation) ---
    % Shows where MPS dies.
    \shade[left color=red!10, right color=red!30, draw=red!40, dashed, thick] 
        (5, 3.5) rectangle (10.5, 7.5);
    \node[align=center, font=\large\bfseries, text=red!70!black] at (7.75, 6.75) {MPS (brutal Truncation)};
    \node[align=center, font=\small, text=red!70!black] at (7.75, 6.25) {\\ \vspace{2pt} \textbf{Hardware Cost (32q): $\sim$4 GPUs}};

        % --- 5. The Hybrid "Hero" Zone ---
        \shade[top color=violet!40, bottom color=violet!60, draw=violet!80, ultra thick, rounded corners=3mm, drop shadow] 
            (4.5, 3.5) rectangle (10.2, 5.7); % slightly taller to fit text
        \node[align=center, font=\small\bfseries, text=white] at (7.35, 4.9) {Hyperion-2 (Partitioned SV-MPS)};
        \node[align=center, font=\small, text=white] at (7.35, 4.1) {locally exact + locally bounded \\ \vspace{2pt} \textbf{Hardware Cost (32q): $\sim$16 GPUs}};
    % --- 6. Arrows explaining the "Accuracy" logic ---
    \draw[->, thick, darkgray] (2.5, 2.8) to[bend left=15] node[above, font=\footnotesize, sloped] {Memory Limits SV} (2.5, 4.5);
    \draw[->, thick, red!80!black] (6, 5.6) to[bend right=15] node[right, font=\footnotesize] {Entanglement Limits MPS} (6, 6.1);

\end{tikzpicture}

    \caption{Conceptual diagram of emulation capabilities as a function of system size and correlation strength}
    \label{fig:emulation_phase_diagram}
\end{figure*}
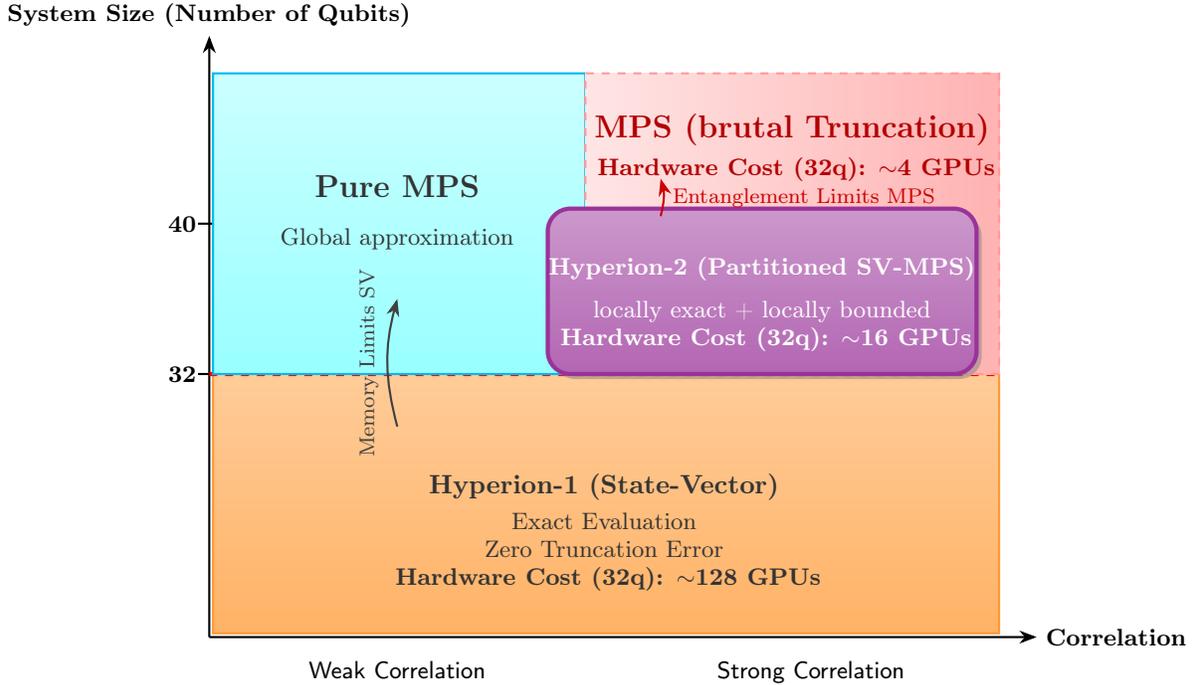

The  advantage of this architecture is twofold: it reduces the memory overhead typically associated with SV approaches when storing the Hamiltonian, second, it enables to minimize numerical errors and accumulation of truncation errors by performing exact evaluations for certain blocks  using exact SV methods, rather than relying only on approximate MPS operations for all evaluations.  We present in Figure \ref{fig:emulation_phase_diagram} 
a conceptual  diagram outlining the  limits of  quantum emulation methods. Sparse SV  methods provide exactness but hit an intractable memory wall around 32 qubits, whereas  MPS emulation scales further but fails  in strongly correlated, highly entangled regimes. By using a locally exact and locally bounded approximation strategy, it extends accurate emulation capabilities up to 40 qubits and reduces the computational footprint, executing 32-qubit workloads on just 16 GPUs compared to the 128 GPUs required by pure state-vector methods.

\subsubsection{Numerical Results}

The proposed emulation  methods MPS and the  Partitioned SV-MPS are evaluated across molecular systems scaling up to 36 qubits. The benchmarked systems include the nitrogen dimer ($N_2$,20 qubits), formic acid using active space approximation ($CH_2O_2$, 28 qubits), and the hydrogen chain ($H_{18}$, 36 qubits), all simulated within the minimal STO-3G basis set (full geometric specifications are provided in the Supplementary Material). All simulations were executed using the Hyperion-2 emulator. For both emulation methods, the molecular Hamiltonian is initially constructed using a  truncation threshold of $10^{-14}$. To prevent memory exhaustion during the massive MPO-MPS contractions inherent to the VQE loop, this Hamiltonian is explicitly represented as a distributed list of MPOs, with the maximum bond dimension of any individual MPO strictly bounded at 100. 

Throughout the optimization, all tensor arithmetic operations that naturally expand the internal bond dimensions are systematically compressed using defined SVD truncation thresholds. Algorithmic performance across the ADAPT-VQE iterations is assessed via three primary metrics: maximum local truncation error, maximum operator gradient magnitude, and ground-state energy convergence.

\begin{figure*}[htb]
    \centering
    \begin{subfigure}[b]{0.48\textwidth}
        \centering
        \includegraphics[width=\textwidth]{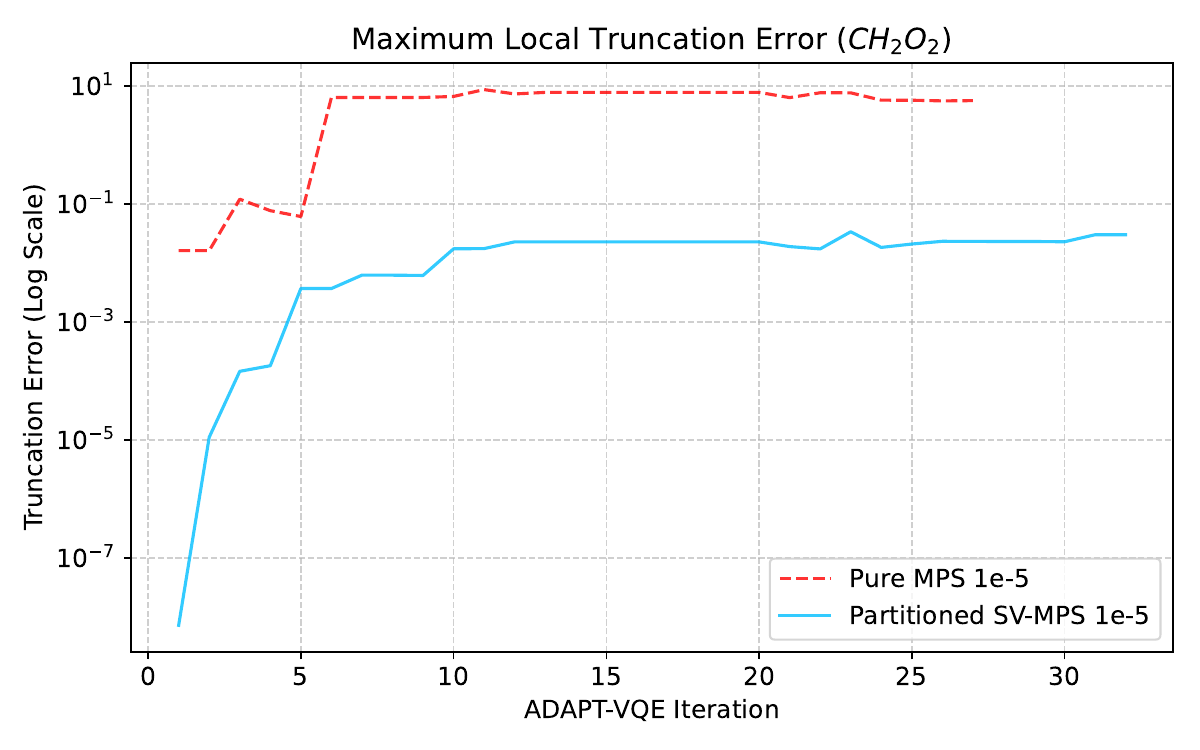}
        \caption{Maximum Local Truncation Error}
        \label{fig:ch2o2_trunc}
    \end{subfigure}
    \hfill
    \begin{subfigure}[b]{0.48\textwidth}
        \centering
        \includegraphics[width=\textwidth]{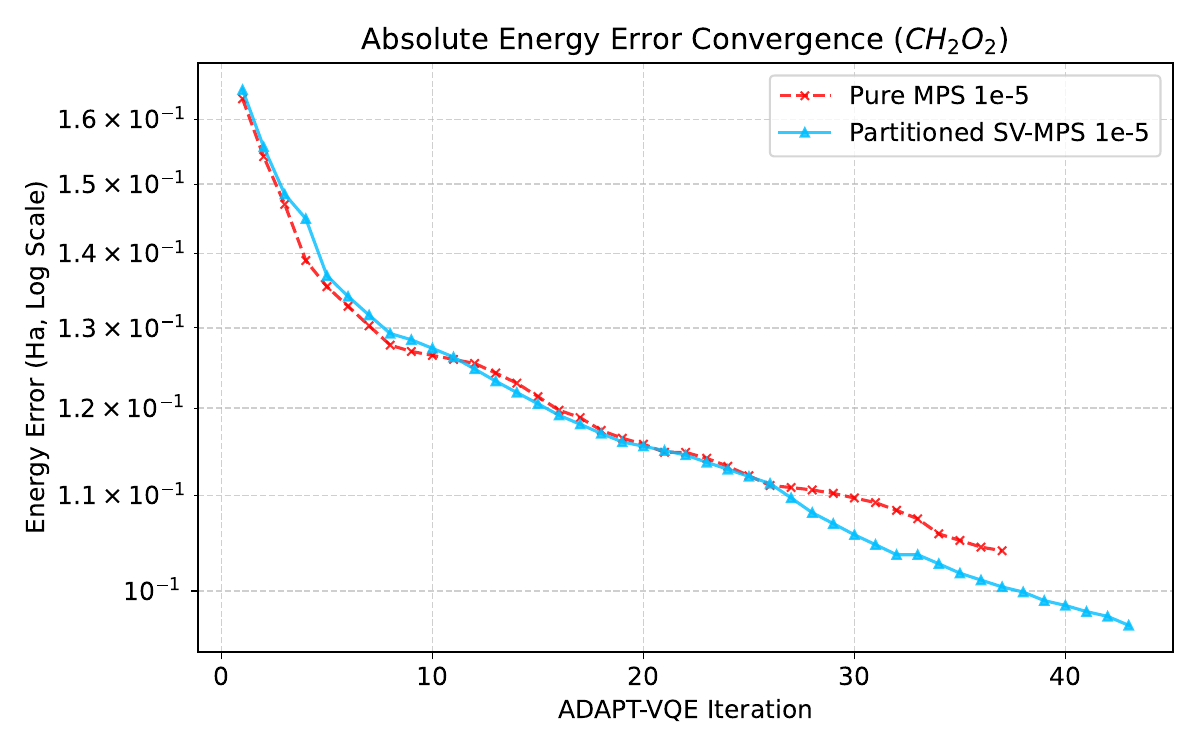}
        \caption{Absolute Energy Error}
        \label{fig:ch2o2_energy}
    \end{subfigure}
    \caption{Performance evaluation of the  $CH_2O_2$ molecule.}
    \label{fig:ch2o2_composite}
\end{figure*}

\begin{figure*}[htb]
    \centering
    \begin{subfigure}[b]{0.48\textwidth}
        \centering
        \includegraphics[width=\textwidth]{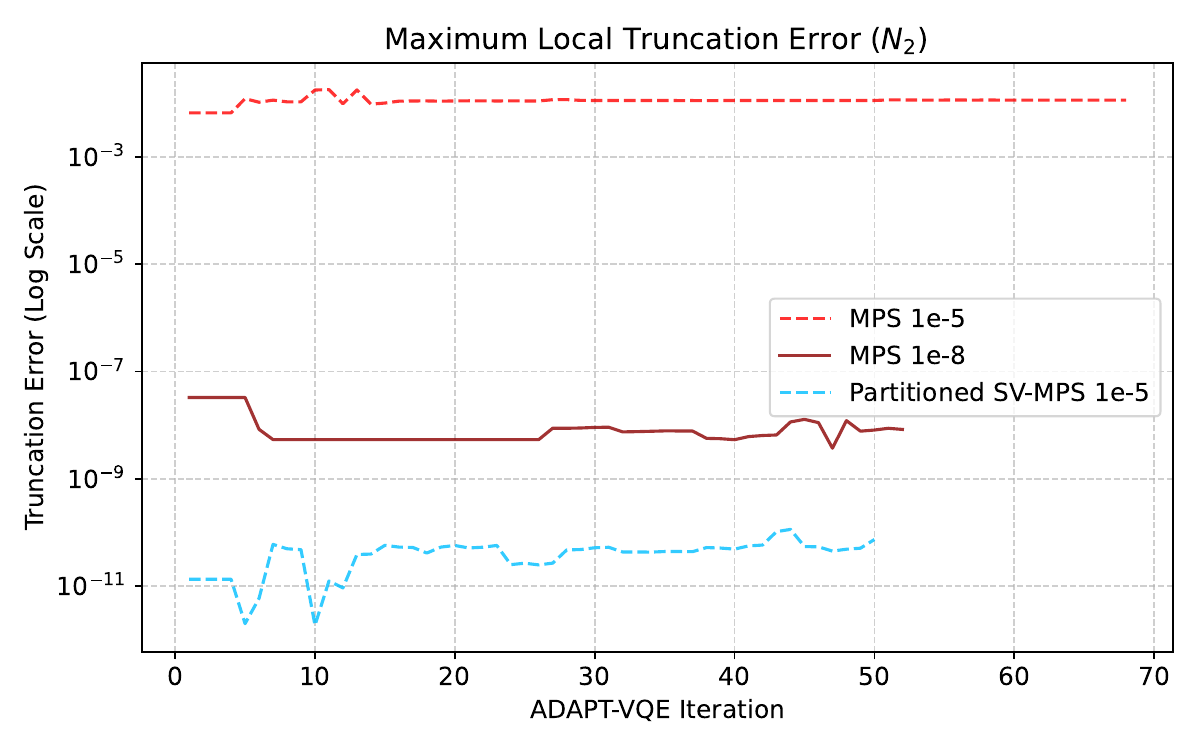}
        \caption{Local Truncation Error}
        \label{fig:n2_trunc}
    \end{subfigure}
    \hfill
    \begin{subfigure}[b]{0.48\textwidth}
        \centering
        \includegraphics[width=\textwidth]{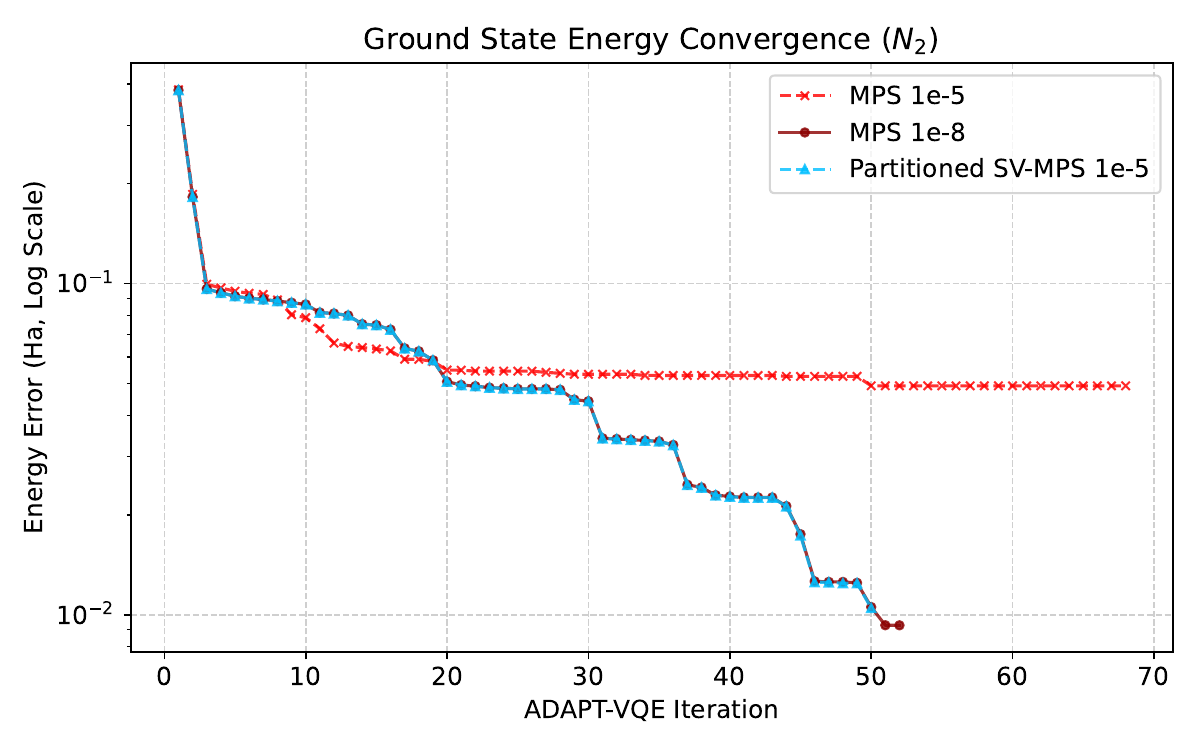}
        \caption{Absolute Energy Error}
        \label{fig:n2_energy}
    \end{subfigure}
    \caption{Performance evaluation of   $N_2$ molecule.}
    \label{fig:n2_composite}
\end{figure*}

To understand some of the limitations of MPS emulation in ADAPT-VQE, Figure~\ref{fig:ch2o2_composite} showcases the cumulative truncation error for the $CH_2O_2$ molecular system. In ADAPT-VQE, evaluating the energy gradient across the vast operator pool necessitates massive MPO-MPS contractions ($\mathbf{H}|\Psi\rangle$). Because applying the Hamiltonian MPO multiplicatively expands the  bond dimension of the updated quantum state, the tensor network must undergo continuous, aggressive SVD compressions to overcome memory overhead. Figure \ref{fig:ch2o2_composite}(a) demonstrates that for pure MPS emulation with a fixed singular-values truncation threshold of $10^{-5}$, this repeated compression  discards critical correlation data, causing the cumulative truncation error to  explode to $\mathcal{O}(1)$. As a direct physical consequence of this compression-induced information loss, Figure \ref{fig:ch2o2_composite}(b) reveals the  MPS expectation value degrading. Conversely, the partitioned SV-MPS approach strictly bounds this error by evaluating the non-interacting Hamiltonian blocks exactly. By exactly evaluating the non-interacting Hamiltonian blocks, this approach shows superior numerical stability under the exact same $10^{-5}$ threshold.  For the case of $N_2$ molecule, see Figure \ref{fig:n2_composite}, the resulting performance mirrors our previous observations: the full MPS approach is incapable of sustaining the entanglement growth demanded by larger ADAPT-VQE iterations. Even when the pure MPS truncation threshold is tightened to $10^{-8}$ (Figure \ref{fig:n2_composite}a), the continuous MPO-MPS compressions generate massive numerical noise.

\begin{figure*}[htb]
        \centering
        \includegraphics[scale=0.4]{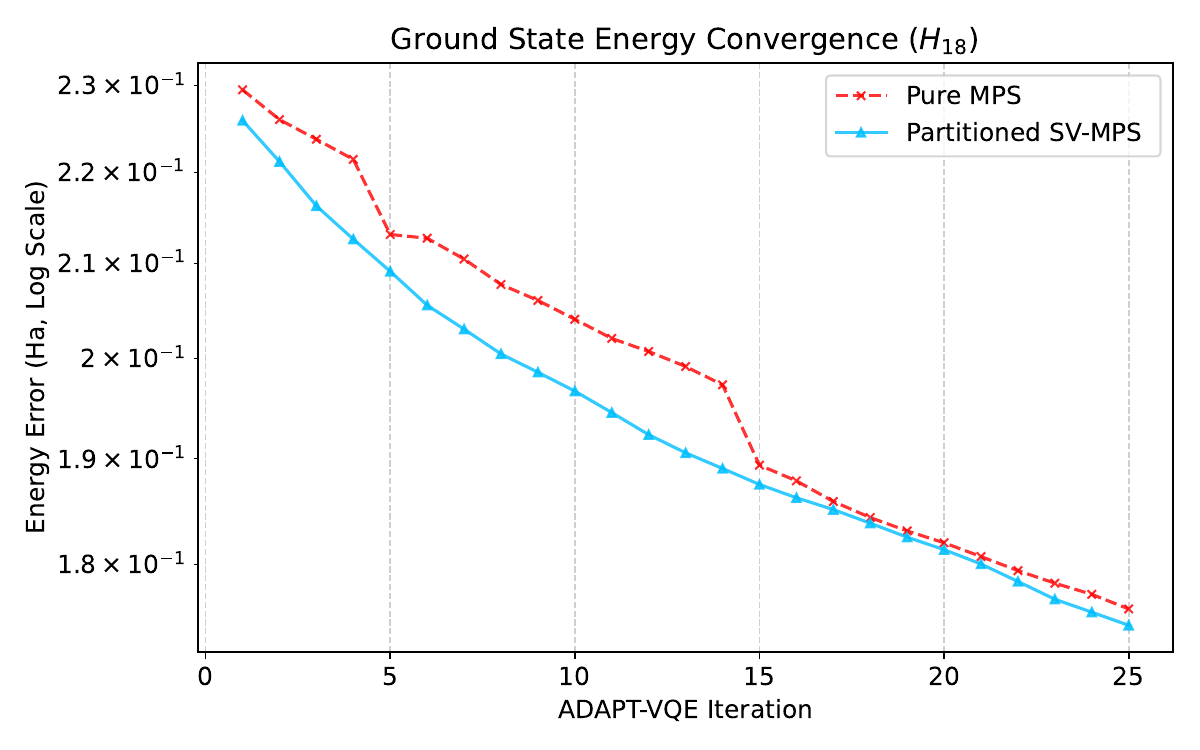}
        \label{fig:h18_energy}
    \caption{Large scale ADAPT-VQE emulation of the 36-qubit $H_{18}$ system.}
    \label{fig:h18_composite}
\end{figure*}

As shown in Figure \ref{fig:h18_composite}, the divergence in the $H_{18}$ energy convergence profiles highlights some of the numerical trade-offs inherent to the MPS method. While MPS can theoretically bypass the 32-qubit memory wall, it does so at the cost of numerical instability. As recently documented in \cite{provazza2024fast}, emulating ADAPT-VQE using MPS introduces coupled errors, arising from the truncated bond dimension, that adds to the inherent errors of the ADAPT convergence scheme. This coupling  obstructs reliable convergence to the exact energy. Consequently, the MPS optimization loop exhibits an unstable energy descent, trapping the system in an elevated small plateau near $2.1 \times 10^{-1}$ Ha. To address these instabilities, we propose the Partitioned SV-MPS method as a robust alternative. Figure \ref{fig:h18_composite},  demonstrates that this partitioned approach achieves superior stability, driving the energy of the 36-qubit system monotonically downward without stagnation all along the 25 iterations.

\begin{table}[htb]
    \centering
    \caption{Simulation capacity limit in logical qubits on NVIDIA H100 (80\,GB) GPUs. The Table compares the initial Hyperion-1 state-vector approach to the new partitioned SV-MPS architecture of Hyperion-2. \textbf{Both implementations can reach CNOT counts beyond 13000.}}
    \label{tab:hyperion-perfs}
    \begin{tabular}{@{} l r c c c @{}}
        \toprule
        & & \multicolumn{3}{c }{\textbf{Maximum Logical Qubits}} \\
        \cmidrule(l){3-5}
        \textbf{H100 GPUs} & \textbf{Total Memory} & \textbf{Hyperion-1} & \textbf{Hyperion-2 (MPS)} & \textbf{Hyperion-2 (SV/MPS partitioned)}\\ 
        \midrule
        1   & \SI{80}{\giga B}    & 24 & 30 & 28 \\
        4   & \SI{320}{\giga B}   & -- & 32 & 30 \\
        8   & \SI{640}{\giga B}   & 28 & -- & 31 \\
        16  & \SI{1280}{\giga B}  & -- & 36* & 32 \\
        64  & \SI{5120}{\giga B}  & -- & -- & 36 \\
        128 & \SI{10240}{\giga B} & 32 & -- & -- \\
        256 & \SI{20480}{\giga B} & -- & 40 & 40** \\
        \bottomrule
  \end{tabular}  
    \footnotesize
    \vspace{0.5cm}

\footnotesize
{$^*$ Although 40 qubits can be reached by the implementation, 36 qubits is the stability limit for the present pure MPS implementation for ADAPT-VQE computations.\\
$^{**}$ 40 qubits is the implementation limit for the partitioned SV/MPS approach (single ADAPT-VQE iteration)} 
\end{table}

Table \ref{tab:hyperion-perfs} benchmarks the maximum logical qubit capacities of Hyperion-1 and Hyperion-2 based on recent ADAPT-VQE simulations executed on NVIDIA H100 (\SI{80}{\giga B}) GPUs. While SV emulation (Hyperion-1) requires 128 GPUs (\SI{10.2}{\tera B} of memory) to simulate a 32-qubit system, the partitioned SV-MPS method evaluates the identical system using only 16 GPUs (\SI{1.2}{\tera B}). This $8\times$ reduction in computational overhead allows Hyperion-2 to successfully extend exact, iterated emulation capacity to 36 logical qubits on 64 GPUs, with a theoretical single-iteration limit of 40 qubits on 256 nodes. This is achieved while preserving
the accuracy of SV methods throughout the ADAPT-VQE iterations, making large-scale quantum
emulation highly accessible on standard HPC clusters.

\section{Conclusion and Perspectives}
Hyperion is a massively parallel, GPU-accelerated quantum emulator designed to address the memory constraints inherent in strongly correlated quantum chemistry simulations. The platform is structured into two primary modules: Hyperion-1, which provides exact state-vector simulations for small-to-medium scale systems, and Hyperion-2, which utilizes Matrix Product States (MPS) for larger-scale applications.The core of Hyperion-2 is a partitioned SV-MPS strategy that hierarchically decomposes the molecular Hamiltonian. This approach directs non-interacting local terms to an exact sparse state-vector core while delegating complex interacting terms to a compressed MPS engine. This partitioned architecture offers several technical advantages. By evaluating non-interacting blocks exactly, the system strictly bounds cumulative truncation errors that often degrade performance in pure MPS emulations during ADAPT-VQE iterations offering enhanced accuracy. The strategy is also hardware efficient as it allows for a 32-qubit simulation using 16 GPUs, representing an 8x reduction in computational overhead compared to the 128 GPUs required for a pure state-vector approach. The platform extends emulation capacity into the 36 to 40 qubit regime for ADAPT-VQE simulations, enabling high-fidelity validation of quantum algorithms for realistic chemical active spaces. 
Beyond chemistry, this general strategy is designed to push the limits of state-vector emulation\cite{de2025universal}, and offers a route to expand the accessible frontier beyond current hardware constraints, i.e. 50 exact qubits.
Future iterations of Hyperion will aim to further reduce computational prefactors and accelerate the convergence of variational algorithms. Hyperion provides a high-fidelity environment for developing quantum algorithms at accuracies approaching the exact Full Configuration Interaction (FCI) and Complete Basis Set (CBS) limits.

\section*{References}
\bibliographystyle{unsrt}
\bibliography{literature}
\end{document}